\documentclass[10pt]{article}
\usepackage{amsmath,amsfonts,amsthm,amssymb,amscd}

\newcommand{\be}{\begin{equation}}
\newcommand{\ee}{\end{equation}}
\newcommand{\bs}{\begin{split}}
\newcommand{\es}{\end{split}}
\newcommand{\ba}{\begin{align}}
\newcommand{\ea}{\end{align}}
\newcommand{\basl}[1]{\begin{align}\begin{split}\label{#1}}
\newcommand{\bas}{\begin{align}\begin{split}}

\newtheorem{theo}{Theorem}[section]
\newtheorem{prop}[theo]{Proposition}
\newtheorem{lemm}[theo]{Lemma}

\newcommand\R{\mathbb{R}}
\newcommand\C{\mathbb{C}}

\title{Semiclassical expansion of the ground state \\ for a model of interacting spins in QED.}

\author{L. Amour and J. Nourrigat}

\date{Universit\'e de Reims, France}

\begin{document}

\maketitle

\begin{abstract}
\noindent
In this article, we consider fixed spin$-1/2$ particles interacting through the quantized electromagnetic field in a constant magnetic field. We give some asymptotic expansions for the ground state and the ground state energy of the  Hamiltonian operator $H(h)$ describing this system.
The first terms of these expansions enable to recover  elementary formulas for the energy and the magnetic field of the spins when considered as
magnets. A first order radiative correction is computed for the energy.

\end{abstract}

\parindent=0pt

\


\


\tableofcontents

\parindent = 0 cm

\parskip 10pt
\baselineskip 17pt

\section{The model.}\label{s1}

The aim of this work is to give semiclassical expansions for the ground state and ground state
energy for  a Hamiltonian operator $H(h)$ modelling the interaction between quantized electromagnetic field and $N$ fixed spin-$1/2$ particles in a constant magnetic field.

 We shall use  a Hamiltonian operator $H(h)$ recalled below in  (\ref{1.14}) and (\ref{1.15})  (see Reuse\cite{REU}, H\"ubner-Spohn \cite{H-S}, Derezi\'nski-G\'erard \cite{D-G}).

 The Hilbert space associated with this Hamiltonian is the completed tensor product ${\cal H}_{ph} \otimes {\cal H}_{sp}$. The Hilbert space  ${\cal H}_{ph}$ for photons may be viewed as the symmetrized Fock space ${\cal F}_s (H _{\C} )$ associated with the complexified of some real Hilbert space $H$ inspired by Lieb-Loss \cite{L-L}. This space $H$  is the space of mappings
 $f = (f_1, f_2, f_3)$ from $\R^3$ to $\R^3$
with $f_j$ belonging in $L^2({\R}^3)$, taking real values and satisfying,
\be\label{1.1}
k_1f_1(k) + k_2f_2(k) + k_3f_3(k) = 0\quad {\rm a.e.}\ee
This space is equipped with the norm,
\be\label{1.2} |f|^2 = \sum_{j=1}^3 \int_{\R^3} |f_j(k)|^2dk. \ee

The Fock space ${\cal F}_s (H _{\C} )$  definition is reminded  in Section \ref{s7}.
The space ${\cal H}_{sp}$ for the spin particles is denoted by $( \C^2  )^{\otimes N}$.

In the space  ${\cal H}_{ph}$, the definition of the model and of the observables  involves three kinds of operators: the number operator $N$, the free photons  energy operator $H_{ph}$ and operators at each point $x\in \R^3$ associated with the three components of the magnetic field. These operators are denoted by $B_m (x)$, $1 \leq m \leq 3$ and for the electric field, it is denoted by  $E_m (x)$, $1 \leq m \leq 3$.
Each of these operators is depending on the semiclassical parameter $h>0$ which is sometimes not explicitly written.

Within the Fock space formalism, the number operator $N$ and the free photons Hamiltonian
 $H_{ph}$  are defined by,
\be\label{1.3}  N= {\rm d}\Gamma (I),\qquad H_{ph} = h {\rm d}\Gamma (M), \ee
$M$ being the multiplication operator by $\omega (k) = |k|$ with domain $D(M) \subset H$, ${\rm d}\Gamma $  is the standard operator (see \cite{RE-SI}) and $h>0$ is the semiclassical parameter. These equalities classically define selfadjoint operators (see \cite{RE-SI}).

In the Fock space formalism, the operators $B_m (x)$ (depending on $h>0$), is defined by,
\be\label{Bm} B_m (x) = \sqrt {h} \Phi_S ( a_m (x) + i b_m (x)) \ee
where, for each $a+i b \in H_{\bf C }$, $\Phi_S (a+i b)$ is the Segal field, defined in \cite{RE-SI}, and $a_m(x)$ and $b_m(x)$ are elements of $H$, therefore  mappings from $\R^3$ into itself, defined by,
\be\label{1.6} a_m (x) (k) =
 {\chi(|k|)|k|^{1\over 2} \over (2\pi)^{3\over 2}} \sin ( k \cdot x )
{k\wedge e_m \over |k|}\ee
\be\label{1.7} b_m (x)   (k) =
 {\chi(|k|)|k|^{1\over 2} \over (2\pi)^{3\over 2}} \cos  ( k \cdot x )
{k\wedge e_m \over |k|}, \ee
where $\chi $ is a function belonging to ${\cal S} (\R)$   and $(e_1, e_2, e_3)$
is the canonical basis of  $\R^3$.

The following estimate will be used:
\be\label{8bis}\Vert  \Phi_S (a+i b)   f \Vert ^2  \leq 2  (|a|^2 + |b|^2)
 \Big [ \Vert    f \Vert ^2 + < Nf , f> \Big ].\ee

Operators in ${\cal H}_{sp}$ use in particular Pauli
 matrices $\sigma_j$ ($1\leq j \leq 3$),
\be\label{1.12} \sigma_1 = \begin{pmatrix}  0 & 1 \\ 1 & 0   \end{pmatrix},
\qquad
   \sigma_2 = \begin{pmatrix} 0 & -i \\ i & 0   \end{pmatrix},
 \qquad
   \sigma_3 = \begin{pmatrix}  1 & 0 \\ 0 & -1  \end{pmatrix}.\ee
For all $\lambda \leq N$ and for any $j\leq 3$,   $\sigma_j^{[\lambda]}$
denotes the following operator in  ${\cal H}_{sp}$,
\be\label{1.13} \sigma_j^{[\lambda]} = I \otimes \cdots \sigma_j \cdots \otimes I,\ee
where $\sigma_j$ is located at the $\lambda ^{th}$ position.

We assume that there are $N$ fixed spin$-1/2$ particles at points $x_{\lambda}$ in $\R^3$
($1\leq \lambda \leq N$).
Denoting the constant magnetic field by $\beta = (\beta_1, \beta_2, \beta _3)$,
the system constituted with these particles and the quantized magnetic field is governed by the operator in ${\cal H}_{ph} \otimes {\cal H}_{sp}$ defined by,
\be\label{1.14} H (h) = H_0 + h H_{int},\qquad H_0 = H_{ph}\otimes I ,\ee
where
\be\label{1.15} H_{int} = \sum _{\lambda = 1}^N \sum_{j=1}^3
( \beta _j + B_j(x_{\lambda}) ) \otimes  \sigma _j ^{[\lambda]}.\ee
It is recalled in \cite{A-L-N-2} (Section 4) that it defines a selfadjoint operator with domain
 $D(H_{ph}) \otimes {\cal H}_{sp}$. In \cite{A-L-N-2} some results of evolution are given, 
 using the pseudodifferential calculus introduced in \cite{A-J-N} and \cite{A-L-N-1}.

It is proved in \cite{B-F-S}, see also \cite{A-H}\cite{G}\cite{H-H-L}\cite{H-S}, that $E_h$, the infimum  of the spectrum   of the operator $H(h)$, is an eigenvalue and the associated eigenspace is of multiplicity $1$.

\begin{theo}\label{t1.2} (i) If $\chi $ is vanishing in a neighborhood of the origin,
and if $\beta \not= 0$, one can find a sequence of real numbers  $\lambda _j$ such that, for all $p$,
\be\label{1.A} \left | E_h - \sum _{j=1}^p \lambda _j h^j \right | \leq C_p h^{p+1}.\ee
(ii) If $\beta \not = 0$, but without the hypothesis on $\chi$, one can find $\lambda _1$ and $\lambda _2$
such that,
\be\label{1.B} |E_h - \lambda _1 h - \lambda _2 h^2 | \leq C h^{5/2}.\ee
\end{theo}

The expansion is formally obtained in Proposition \ref{p7.3} and the control of the remainder term is derived in Theorem \ref{t7.7}. Point $(ii)$ is proved at the end of Section \ref{s7}. For the first two terms, one finds,
$$ \lambda _1 h +  \lambda _2 h^2 =  - N h |\beta | - {h^2 \over 2}
 \sum _{\lambda ,\mu \leq N } F ( x_{\lambda } - x_{\mu}) - NC h^2,$$
where $F$ is the semiclassical interaction function between parallel spins, given by, when the spins are aligned along the direction ${\bf n}_{\beta} = \beta /|\beta|$,
\be\label{6.6} F(x) = (2 \pi )^{-3}  \int _{\R^3 } |\chi(|k|) |^2
  \cos ( k \cdot x) \ \left ( 1 -  \frac { |k\cdot {\bf n}_{\beta} |^2}  { |k|^2} \right ) 
  dk \ee
and $C$ is defined in (\ref{7.2.28}). The first term amounts to the sum of the energies of each spin aligned along the direction of the constant field $\beta$. 
 The second term amounts to the sum of the classical interaction energies between two spins, all being parallel (including for the auto-interaction). See the comments after Proposition \ref{p7.5}.  
Only the third term $NC h^2$ is genuinely a quantum term.

Now assuming that $\chi $ is vanishing in a neighborhood of the origin, it is proved in Theorem \ref{t7.2} that a unitary eigenvector $\varphi_h$ has, up to a normalization factor, an asymptotic expansion in powers of  $h^{1/2}$,  to any order.  Without this hypothesis, we can
give only an expansion with only three terms.

For all $x\in \R^3$, we can compare the average magnetic field $< {\bf B} (x)  \varphi _h , \varphi _h >$ taken on the ground state $\varphi_h $ with the magnetic field $ {\bf B} ^{class } (x) $
associated by elementary physics with the spins systems regarded as magnets all being aligned along the direction of the (non zero) constant  magnetic field.
Setting ${\bf n }_{\beta } = \beta / |\beta |$, we have to consider the current density ${\bf j } (x) $ corresponding to this spins system,
\be\label{current} {\bf j } (x) = h  {\bf n }_{\beta } \wedge {\rm grad} \Phi (x), \ee
\be\label{1.25}  \Phi (x) = \sum _{\lambda = 1}^N \rho (x - x_{\lambda }),\qquad
\rho (x) = (2 \pi )^{-3}  \int _{\R^3} |\chi (k)|^2  \ \cos (k \cdot  x) dk.\ee
The potential vector $ {\bf A} ^{class } (x) $ satisfies $\Delta  {\bf A} ^{class }  =
{\bf j }$ and also,
$$ {\bf A} ^{class } (x) = {1\over 4 \pi} \int _{\R^3} { {\bf j } (y)\over |x-y|} dy. $$
The classical magnetic field is,
  $$  {\bf B} ^{class }= {\rm rot } {\bf A} ^{class }  $$
and the electric field  ${\bf E} ^{class }$ is zero. One notes the role of the function  $\chi$: the spin particle is not exactly a point-like particle.

We shall prove the following theorem.

\begin{theo}\label{t1.3} Suppose that $\chi (0) = 0$ and $\beta \not= 0$. For all $x$ in $\R^3$, we have
$$ < {\bf B} (x)  \varphi _h , \varphi _h >    -  {\bf B} ^{class }(x) = {\cal O } (h^{3/2} ), $$
$$ < {\bf E} (x)  \varphi _h , \varphi _h >    -  {\bf E} ^{class }(x) = {\cal O } (h^{3/2} ).$$
\end{theo}

The first step of the proof is Proposition \ref{p7.6}. The second step appears at the end of Section \ref{s7}. The proof only used the fact that $\chi (0) = 0$ and not that $\chi $ is vanishing in a neighborhood of the origin. Nevertheless, the method used for these estimates, which relies on a conjugate operator, does not allow to avoid the hypothesis $\chi (0) = 0$.

\section{Asymptotic expansions for the ground state.}\label{s7}

The following theorem is proved in \cite{B-F-S}. Let $b_0$ and $b_1$ be unitary elements of ${\bf C}^2$ such that:
$$ \sum _{m=1}^3 \beta _m \sigma_m b_{0 } = - |\beta | b_{0 }
 \hskip 2cm
 \sum _{m=1}^3 \beta _m \sigma_m b_{1 } =  |\beta | b_{1 }$$
For all $E \subset \{ 1 , \dots , N \}$,  $a_E$ denotes the following element,
\be\label{6.4} a_E = a_1 \otimes \cdots \otimes a_N, \qquad  a_j = \left \{ \begin{matrix}  b_1 & {\rm if} &j\in E \\
 b_0 & {\rm if} & j \notin E  \end{matrix}\right ..\ee

\begin{theo}\label{t7.1}
(\cite{A-H}\cite{B-F-S}\cite{G}\cite{H-H-L}\cite{H-S})
The infimum $E_h$ of the spectrum of $H(h)$
satisfies
\be\label{7.1}E_h \leq -N |\beta | h.\ee

The eigenspace associated with $E_h$ has dimension $1$.
Moreover, there exists an unitary eigenvector $\varphi_h$ corresponding to the eigenvalue $E_h$, the infimum of the spectrum of $H(h)$, such that, for any small enough $h$,
\be\label{7.2}\Vert  \varphi_h -  ( \Psi_0 \otimes a_{\emptyset}) \Vert \
\leq C h ^{1/2}.\ee
\end{theo}

The estimate (\ref{7.1}) follows from
$ < H(h) ( \Psi_0 \otimes a_{\emptyset }), ( \Psi_0 \otimes a_{\emptyset }) > = -N |\beta | h$. The estimate (\ref{7.2}) is also a consequence of
Proposition \ref{p7.8}  below. The constant $C$ coming from this  Proposition could may be different from the one in \cite{B-F-S}. When Proposition \ref{p7.8}  is used, the constant $C$ depends on the $L^2(\R^3)$ norms of the functions $\chi (|k|) |k|^{1/2}$ and
$\chi (|k|) |k|^{-1/2}$.

The aim of this section is to establish an asymptotic expansion as $h$ tends to $0$ of the eigenvalue $E_h$ and of an unitary eigenvector $\varphi_h$.

\subsection{Statement.}\label{s7.1}

\begin{theo}\label{t7.2}
Suppose that there exists $\rho >0$ such that the function $\chi$ in (\ref{1.6}) and (\ref{1.7}) vanishes for $|k|\leq \rho $.  Assume that $\beta \not = 0$. Let $E_h$ be the infimum of the spectrum of  $H(h)$ and let $\varphi_h$ be a corresponding normalized eigenvector of $H(h)$. Then, there exists a sequence of elements of  ${\cal H}_{ph} \otimes {\cal H}_{sp} $  denoted
 $u_{j}$ ($j\geq 0$) and a sequence of real numbers $\lambda _j$ ($j\geq 1$) such that,

 (i) One has,
\be\label{7.3} u_0 = \Psi_{0} \otimes a_{\emptyset},\qquad \lambda _1 = - N |\beta |.\ee
 (ii)  For all integers $p$, there exists $C_p$ satisfying for  $h$ small enough,
\be\label{7.4} \left | E_h - \sum_{j= 0} ^p \lambda _j h^j  \right | \leq  C h^{p +1}.\ee
 (iii)  Moreover, for all integers $p$ and any $h >0$, there exists $\rho_h >0$ and $\theta _h $
 (depending on $h$ and $m$) such that,  setting
\be\label{7.5}  V_{2p+1} (h) = \sum_{j= 0} ^{2p+1}  u _j h^{j/2}  - \rho_h e^{i\theta _h } \varphi_h,\ee
we have for $h$ sufficiently small,
\be\label{7.6} \Vert  V_{2p+1} (h)  \Vert \leq C_p h^{p+1},\ee
where the constant $C_p$ is independent on $h$.

(iv) We also have,
\be\label{7.7}<  (N  \otimes I)  V_{2p+1} (h) ,  V_{2p+1} (h)  >\  \leq C_p h^{2p+1}.\ee
\end{theo}

\subsection{Formal construction of the expansion.}\label{s7.2}

{\it Additional details on the Fock space.}  Let us recall that,
\be\label{7.2.1} {\cal F}_s (H_{\C}) =  \oplus _{m\geq 0} {\cal F}_m, \ee
where ${\cal F}_0 = {\C}$ and ${\cal F}_m$ is completion of the $m-$fold symmetric tensor product   $H_{\C} \odot \cdots \odot H_{\C} $. One may
then consider an element of ${\cal F}_m$ as a symmetric map  $f$ from
$ (\R^3)^m $ to $ (\C^3) ^{\otimes m}$ satisfying for all $a_2,\dots,a_m$ in $\{ 1, 2, 3 \}$ and for all $k_1,\dots,k_m$ in $\R^3$,
\be\label{7.2.2} \sum _{j=1}^3 { k_{1 , j}}  f_{ j, a_2 , ... a_m} ( k_1 , \dots , k_m) = 0.\ee
We use here the notation $k_1 = (  k_{1 , 1} ,  k_{1 , 2} ,  k_{1 , 3})$. In addition, the components of this function $f$ should be in $L^2 (\R^{3m})$, which is  defining the norm in ${\cal F}_m$.
Thus, an element of $ {\cal F}_s (H_{\C})$ is a sequence $ f = (f_m)_{(m\geq 0 )}$ where $f_m$ is an
element of ${\cal F}_m$ and one has,
\be\label{7.2.3} \Vert f \Vert ^2 = \sum _{m\geq 0} \Vert f_m \Vert ^2.\ee

We shall denote by $\Psi_0$ a unitary element of ${\cal F}_0$.

For any $\rho >0$ and $m\geq 1$,  ${\cal F}_m (\rho )$ stands for the set of elements
$f$ in  ${\cal F}_m  $ satisfying $f(k_1 , ... , k_m) $ belongs to ${\cal S}(\R^{3m})$ and is vanishing if one of the $|k_j|$ is
$\leq \rho $. If $m=0$, it is agreed that ${\cal F}_0 (\rho )= {\cal F}_0 $. It is also agreed
that ${\cal F}_m   =0$ if $m<0$. One sets,
\be\label{7.2.4} {\cal F} _{even} (\rho) =  {\cal F}_0 \oplus {\cal F}_2 (\rho) \oplus {\cal F}_4 (\rho) \oplus \cdots\ee
\be\label{7.2.5} {\cal F} _{odd} (\rho) =  {\cal F}_1 (\rho)  \oplus {\cal F}_3 (\rho) \oplus {\cal F}_5 (\rho) \oplus \cdots.
\ee
Elements in these spaces here are finite sums.

Let us recall that, if $M: H \rightarrow H$ is the multiplication by
$\omega (k) = |k|$ and if $f$ is a rapidly decreasing function in ${\cal F}_m$
then one has,
\be\label{7.2.6} ( {\rm d}\Gamma (M) f ) ( k_1 , \dots , k_m) = ( |k_1| + \dots + |k_m | ) f  ( k_1 , \dots , k_m).\ee

We remind that $H_{ph} = h {\rm d}\Gamma (M)$.
For all $x\in \R^3$ and for each $m\leq 3$, the operator $B_m (x)$ corresponding to the semiclassical parameter $h$ is defined by (\ref{Bm}). Therefore, we can write,
\be\label{7.2.16} H(h) =   h K_1 + h^{3/2} K_{3/2},\ee
with
\be\label{7.2.17}K_1 = {\rm d}\Gamma (M) \otimes I + I \otimes T_0,\quad
T_0 =  \sum _{\lambda = 1}^N \sum _{m=1}^3  \beta_m  \sigma _m^{[\lambda ]},\ee
\be\label{7.2.18} K_{3/2} = \sum _{\lambda = 1}^N \sum _{m=1}^3 \Phi_S ( a_m( x_{\lambda }) + i
b_m( x_{\lambda }) ) \otimes  \sigma _m^{[\lambda ]},
\ee
where $a_m( x_{\lambda })$ and $b_m( x_{\lambda })$ are defined in (\ref{1.6}) and  (\ref{1.7}). The operators $ K_{1}$
 and $ K_{3/2}$ are independent on $h$.

With these notations, the formal construction of the asymptotic expansion is provided by the following Proposition.

\begin{prop}\label{p7.3} Let $\rho >0$ be such that the function $\chi$ in (\ref{1.6})(\ref{1.7})  is vanishing for
 $|k|\leq \rho $. Suppose $\beta \not = 0$.
Then, there exists a sequence of elements in  ${\cal H}_{ph} \otimes {\cal H}_{sp} $ denoted by
 $u_{j}$ ($j\geq 0$) and a sequence of real numbers $\lambda _j$ ($j\geq 1$) such that,
 $u_0$ and  $\lambda _1$ are given in (\ref{7.3}) and if $j$ is even,
\be\label{7.2.8}  u_j \in {\cal F}_{even}^{(j)} (\rho)  \otimes {\cal H}_{sp},\ee
 and if $j$ is odd,
\be\label{7.2.9}  u_j \in {\cal F}_{odd }^{(j)} (\rho)   \otimes {\cal H}_{sp} \ee
and such that, for all integers $p$, setting,
\be\label{7.2.10}  U^{(p)} (h) = \sum _{j= 0}^p u_j h^{j/2},
\qquad \lambda ^{(p)} (h) = \lambda _1 h + \cdots + \lambda _{p} h^{p},\ee
we have
\be\label{7.2.11}  \Big (H(h) - \lambda ^{(p+1)} (h) \Big )  U^{(2p)} (h)= R^{(2p)} (h),\ee
\be\label{7.2.12} \Big (H(h) - \lambda ^{(p+1)} (h) \Big )  U^{(2p+1)} (h)= R^{(2p+1)} (h),\ee
where the $R^{(j)} (h)$ are expressed as following,
\be\label{7.2.13}  R^{(2p)} (h) = \sum_{k\geq 1}   h^{ p + 1 + (k/2)} f_{2p }^{ (p + 1 + (k/2)) },
\ee
with the $f_{2p }^{ (p + 1 + (k/2)) }$ being elements of   ${\cal F }_s ( H _{\C} ) \otimes {\cal H}_{sp}$,
\be\label{7.2.14}  R^{(2p+1)} (h)  = \sum_{k\geq 0 } h^{ p + 2 + (k/2)}
 f_{2p +1 }^{ (p + 2 + (k/2)) },\ee
with the $ f_{2p +1 }^{ (p + 2 + (k/2)) } $ belonging to ${\cal F }_s ( H _{\C} )\otimes {\cal H}_{sp}$.
The sums in the right hand sides of  (\ref{7.2.13}) and (\ref{7.2.14}) are finite.
\end{prop}

The above elements  $u_{j}$ are independent on  $h$. The proof uses the following Lemma.

\begin{lemm}\label{l7.4}
Let $\lambda _1$  be defined in  (\ref{7.3}), and $T_0$ in (\ref{7.2.17}). 
Then, for all  $f$ in  ${\cal F} _{odd} (\rho)\otimes {\cal H} _{sp}$,
 (resp. in ${\cal F} _{even} (\rho)\otimes {\cal H} _{sp}$), there exists
 $u$ in ${\cal F} _{odd} (\rho)\otimes {\cal H} _{sp}$,
 (resp. in ${\cal F} _{even} (\rho)\otimes {\cal H} _{sp}$) satisfying,
\be\label{7.2.15} ( {\rm d}\Gamma (M) \otimes I   +  I \otimes (T_0  -  \lambda _1)  ) u  =  f - \Pi f,\ee
where $\Pi $ is the orthogonal projection on
  $u_0 = \Psi_0 \otimes a_{\emptyset}$.
\end{lemm}

Note that $\Pi f  = 0$ if $f$ is in ${\cal F} _{odd} (\rho)\otimes {\cal H} _{sp}$.

{\it Proof of the Lemma.}  One may write,
$$ f = \sum _{E \subset \{ 1, \dots , N \} } \sum _{m\geq 0}  f_{E, m }
\otimes   a_E, $$
with $f_{E, m } $ in  ${\cal F} _{m} (\rho) $, vanishing for even $m$  (resp. for odd $m$
). If $ m\geq 1$, set
$$  u_{E, m } (k_1 , \dots , k_m) = { f_{E, m } (k_1 , \dots , k_m) \over |k_1| + \cdots + |k_m| + 2 |\beta | |E| }.$$
Since $f_{E, m } $ vanishes in neighborhood of the origin, this element is well defined, even when $E$ is empty. If $m= 0$ and $E \not = \emptyset$, set
$$  u_{E, 0 }  = {  f_{E, 0 } \over  2 |\beta | |E| }. $$
Then set,
$$ u =  \sum _{m\geq 1}  \sum _{E \subset \{ 1, \dots , N \} }  u_{E, m }
\otimes   a_E  +  \sum _{E \subset \{ 1, \dots , N \}\atop E \not = \emptyset  }  u_{E, 0 }
\otimes   a_E.  $$
This element $u$ has the stated properties in the Lemma.

\hfill $\Box$

 {\it Proof of Proposition \ref{p7.3}.  } Note that the operator $K_1$ defined in (\ref{7.2.17}) maps each of the two spaces
${\cal F }_{even } (\rho) \otimes {\cal H } _{ph}$  and ${\cal F }_{odd } (\rho) \otimes {\cal H } _{ph}$
into itself, whereas $K_{3/2}$ defined in (\ref{7.2.18}) maps each of these two spaces into each other. This comes from the Segal field $\Phi_S $  definition (see \cite{RE-SI}).
By writing that the coefficient of $h^j$ ($j\leq p+1$) in the left hand side of (\ref{7.2.11}) or (\ref{7.2.12}) is zero, one sees that the $u_j$ and the $\lambda _j$ have to satisfy the following relations,
\be\label{7.2.19}( K_1 - \lambda _1 ) u_0 = 0,\ee
\be\label{7.2.20} ( K_1 - \lambda _1 ) u_1 + K_{3/2} u_0 = 0,\ee
\be\label{7.2.21} ( K_1 - \lambda _1 ) u_2 + K_{3/2} u_1 - \lambda _2 u_0 = 0,\ee
\be\label{7.2.22} ( K_1 - \lambda _1 ) u_3 + K_{3/2} u_2 - \lambda _2 u_1 = 0.\ee
More generally, if $m= 2p$ is even, one needs,
\be\label{7.2.23}( K_1 - \lambda _1 ) u_{2p} + K_{3/2} u_{2p-1}  - \lambda _2 u_{2p-2} - ... -
  \lambda _{p +1}  u_{0} = 0\ee
and if $m= 2p+1$ is odd,
\be\label{7.2.24} ( K_1 - \lambda _1 ) u_{2p+1} + K_{3/2} u_{2p}  - \lambda _2 u_{2p-1} - ... -
  \lambda _{p+1}  u_{1} = 0.\ee
One has,
 $ {\rm d}\Gamma (M) \Psi_0 = 0$ and $(T_0 - \lambda _1) a_{\emptyset } = 0 $, thus,
 the elements $u_0$ and $\lambda _1$ defined in  (\ref{7.3}) satisfy (\ref{7.2.19}).
Since the operator $K_{3/2}$ exchanges parity, $K_{3/2} u_0$ is in ${\cal F}_{odd } (\rho)   \otimes {\cal H}_{sp}$. According to the
 Lemma \ref{l7.4}, there exists $u_1$ in ${\cal F}_{odd } (\rho)   \otimes {\cal H}_{sp}$
satisfying (\ref{7.2.20}). Set $p\geq 0$. Suppose that $u_0,\dots,u_{2p+1}$ and  $\lambda _1,\dots,\lambda _{p+1}$ satisfying  (\ref{7.2.23}) and (\ref{7.2.24}) are already constructed.
In order to determine $u_{2p+2}$ and $\lambda _{p+2}$, one applies Lemma \ref{l7.4}
with,
\be\label{7.2.25} f _{2p+2} = \left \{
\begin{array}{lll}
- K_{3/2} u_{2p+1} + \lambda _2 u_{2p} + \cdots + \lambda _{p+1} u_2
 & {\rm if} & p\geq 1 \cr \cr
 - K_{3/2} u_{1} \hfill & {\rm if} & p= 0  \cr
 \end{array}
 \right .. \ee
 Since $K_{3/2}$ exchanges parity, this element belongs to
 ${\cal F}_{even } (\rho)   \otimes {\cal H}_{sp}$. One defines
 $\lambda _{p+2}$ by,
\be\label{7.2.26}  \lambda _{p+2} =- < f _{2p+2} , u_0 >.\ee
 According to Lemma \ref{l7.4}, there exists $u_{2p+2}$ in ${\cal F}_{even } (\rho)   \otimes {\cal H}_{sp}$
  such that,
$$ ( K_1 - \lambda _1 ) u_{2p+2} =  f _{2p+2} +  \lambda _{p+2} u_0, $$
that is to say, (\ref{7.2.23}) with $p$ replaced by $p+1$. To get $u_{2p+3}$,  Lemma \ref{l7.4} is applied with,
 $$ f _{2p+3} = - K_{3/2} u_{2p+2} + \lambda _2 u_{2p+1} + ... + \lambda _{p+2} u_1. $$
 This element belongs to ${\cal F}_{odd } (\rho)   \otimes {\cal H}_{sp}$ and consequently,
 $\Pi  f _{2p+3} = 0$. According to Lemma \ref{l7.4}, there indeed exists  $u_{2p+3}$ in
 ${\cal F}_{odd} (\rho)   \otimes {\cal H}_{sp}$ satisfying
$$ ( K_1 - \lambda _1 ) u_{2p+3} =  f _{2p+3}. $$
We have therefore constructed the sequences  $(u_j)$ and $(\lambda _j) $ satisfying  (\ref{7.2.23}) and (\ref{7.2.24}). The properties in the statement of the  Proposition then follows.

\hfill $\Box$

The elements $u_0$ and $\lambda _1$ are defined in  (\ref{7.3}). The following Proposition gives an explicit computation of
$u_1$ and $\lambda _2$. One sees that here, we do not  need
any  hypothesis on the behaviour of $\chi $ in a neighborhood of the origin.

\begin{prop}\label{p7.5} We have,
\be\label{7.2.27} \lambda _2 = - N C - {1\over 2} \sum _{\lambda ,\mu \leq N } F ( x_{\lambda } - x_{\mu}),\ee
where $F$ is the semiclassical parallel spins interaction function defined in  (\ref{6.6}) and
\be\label{7.2.28} C ={1\over 2}  (2\pi)^{-3} \int _{\R^3} { \chi (|k|)^2 |k| \over |k| + 2 |\beta |} \
{ |k|^2 + k_3^2 \over |k|^2 } dk.\ee
\end{prop}

{\it Proof.}  Let us first precise the computation of $u_1$. We assume that $\beta = (0, 0, |\beta|)$. 
According to (\ref{7.2.18}),
$$ K_{3/2} u_0  = f_{\emptyset} \otimes a_{\emptyset} + \sum _{\mu = 1}^N
f_{\mu} \otimes a _{ \{ \mu \} }. $$
Since $\sigma_3^{[\mu ]} a_{\emptyset } = - a_{\emptyset }$, we have,
\be\label{7.2.29} f_{\emptyset} = - \sum _{ \mu \leq N } \Phi_S ( a_3( x_{\mu }) + i
b_3( x_{\mu }) )  \Psi_0.\ee
Similarly, since $\sigma_1^{[\mu ]} a_{\emptyset } =  a_{\{ \mu \} }$ and $\sigma_2^{[\mu ]} a_{\emptyset } = - i a_{\{ \mu \} }$,
\be\label{7.2.30} f_{\mu} =  \Phi_S ( a_1( x_{\mu }) + i
b_1( x_{\mu }) )\Psi_0  - i \Phi_S ( a_2( x_{\mu }) + i
b_2( x_{\mu }) )\Psi_0.\ee
All these elements are in ${\cal F}_1 = H _{\C}$. For all
$X$ in $H _{\C}$, we can identify $\Phi_S (X) \Psi_0$ with $X/ \sqrt {2}$
which is therefore a function of $k\in \R^3$.
The element $u_1$ needs to satisfy (\ref{7.2.20}). In view of Lemma \ref{l7.4}, it can be written as,
\be\label{7.2.31} u_1 =  u_{\emptyset} \otimes a_{\emptyset} + \sum _{\mu = 1}^N
u_{\mu} \otimes a _{ \{ \mu \} },\ee
where $u_{\emptyset}$ and the $u_{\mu}$ are in ${\cal F}_1$, defined by,
\be\label{7.2.32} u_{\emptyset} (k) = - { f_{\emptyset} (k) \over |k|},\qquad
 u_{\lambda} (k) = -  { f_{\lambda} (k) \over |k| + 2 |\beta |}.\ee
According to (\ref{7.2.26}) and (\ref{7.2.25}) (with $p=0$), we have,
\be\label{7.2.33} \lambda _2 =  < K_{3/2} u_{1} ,  u_0 > = <  u_{1} , K_{3/2} u_0 >.\ee
Consequently,
$$ \lambda _2 = <u_{\emptyset} , f_{\emptyset} >  + \sum _{\lambda  = 1}^N
< u_{\lambda} , f_{\lambda} >. $$
We have,
$$ <u_{\emptyset} , f_{\emptyset} >  =- {1\over 2}  \sum _{\lambda , \mu \leq N}
\int _{\R^3} \Big (  a_3( x_{\lambda }) + i
b_3( x_{\lambda  })\Big ) (k) \cdot  \Big (  a_3( x_{\mu }) - i
b_3( x_{\mu })\Big ) (k) { dk \over |k| }.   $$
Therefore, using (\ref{1.6}) and (\ref{1.7}),
$$ <u_{\emptyset} , f_{\emptyset} >   = -  {1\over 2}  \sum _{\lambda , \mu \leq N}
 F (x_{\lambda } -x_{\mu } ),   $$
where $F$ is the semiclassical parallel spins interaction function defined in  (\ref{6.6}). We similarly see that,
$$ < u_{\lambda} , f_{\lambda} > =-  {1\over 2}  \int _{\R^3}
  \Big | ( a_1( x_{\lambda }) + i b_1( x_{\lambda }) ) - i
  ( a_2( x_{\lambda }) + i b_2( x_{\lambda }) \Big | ^2
  { dk \over |k| + 2 |\beta | }. $$
One again uses (\ref{1.6}) and (\ref{1.7}) noticing that,
$ | ( k \wedge e_1) - i ( k \wedge e_2) |^2 = |k|^2 + k_3^2$.
 Consequently,
$$ < u_{\lambda} , f_{\lambda} > = -  C
 $$
where $C$ is defined in (\ref{7.2.28}).

\hfill $\Box$

For the interpretation of the function $F$, we note that the classical potential vector associated to the current density
 ${\bf J} (x) = {\bf n}_{\beta} \wedge {\rm grad} \rho (x)$,
where $\rho $ is defined in (\ref{1.25}), is
 $$ {\bf A} ^{class } (x) = (2 \pi )^{-3}  \int _{\R^3 } |\chi(|k|) |^2
  \sin ( k \cdot x ) \   { k \wedge {\bf n}_{\beta}  \over |k|^2 } dk. $$
The magnetic field is ${\bf B} ^{class }  = {\rm rot} {\bf A} ^{class }$.
 In particular, its projection on the direction of $\beta$  is $F(x)$, with  $F$ defined in (\ref{6.6}).
By translation, $ F(x _{\lambda }  - x_{\mu } )$ is the classical magnetic field created by the spin centered at  $x_{\lambda}$ (that is to say, by the current density
 $ {\bf n}_{\beta} \wedge {\rm grad} \rho ( x - x_{\lambda }) $), taken at  $x_{\mu}$ and projected on the direction where all the spins are aligned (generated by ${\bf n}_{\beta}$).
 According to the coupling constants, one can think that
  $  - h^2  F(x _{\lambda }  - x_{\mu } )$ is
  the interaction energy of the spins centered at
  $x_{\lambda }$ and $x_{\mu}$, aligned and pointing in the same direction parallel to
  ${\bf n}_{\beta}$.

We shall now compute the average magnetic field taken on the ground state first order asymptotic expansion. We do not have any  hypothesis on the behaviour of $\chi $ at the origin.

\begin{prop}\label{p7.6}  We have, for any $x\in \R^3$,
\be\label{7.2.34} < ({\bf B}  (x)\otimes I)  ( u_0 + h^{1/2} u_{1} ), ( u_0 + h^{1/2} u_{1} ) >  =
{\bf B}^{class}   (x),\ee
\be\label{7.2.35} < ({\bf E}  (x)\otimes I)  ( u_0 + h^{1/2} u_{1} ), ( u_0 + h^{1/2} u_{1} ) >  =
0,\ee
where ${\bf B}^{class}   (x)$ is defined in  Section \ref{s1}.
\end{prop}

{\it Proof.}  One can suppose that $\beta = (0, 0 , |\beta|)$. The above computations  show that the classical magnetic associated to the current density defined in (\ref{current}) with  $n_{\beta} = e_3 = (0, 0, 1)$,
is
\be\label{7.2.36} B_m ^{class} (x) = h  (2 \pi )^{-3} \sum _{\lambda = 1}^N \int _{\R^3} |\chi (k)|^2
\ \cos (k \cdot ( x - x_{\lambda } )) \ { ( k \wedge e_m ) \cdot ( k \wedge e_3 )\over |k|^2 } dk.\ee
Besides, we have,
$$ <( B_m (x)\otimes I) ( u_0 + h^{1/2} u_{1} ), ( u_0 + h^{1/2} u_{1} ) >  =  2  h^{1/2}  {\rm Re}
< (B_m (x)\otimes I)   u_0 , u_{1} >.   $$
Indeed, for all $u$ belonging to one of the  ${\cal F}_j$, we have $< B_m(x) u , u > = 0$.
According to the construction (\ref{7.2.31}) and (\ref{7.2.32})
of $u_1$ in the proof of Proposition \ref{p7.5}, we have,
$$ < (B_m (x)\otimes I)   u_0 , u_{1} >  = < B_m (x) \Psi_0  , u_{\emptyset} >. $$
Using the expression (\ref{7.2.32}) of $ u_{\emptyset} $ and next, the one of $ f_{\emptyset} $ in (\ref{7.2.29}),
both considered as  elements of $H_{\C}$, we obtain,
$$ < (B_m (x)\otimes I)   u_0 , u_{1} >  = \int _{\R^3} \Big ( a_m (x) + i b_m (x) \Big )
(k) \cdot { \overline {f_{\emptyset}  (k) } \over |k|} dk $$
$$ = \sum _{ \mu \leq N }  \int _{\R^3} \Big ( a_m (x) + i b_m (x) \Big )
(k) \cdot { ( a_3( x_{\mu }) - i b_3( x_{\mu }) ) (k) \over |k|} dk.   $$
One therefore recovers the right hand side of  (\ref{7.2.36}).

\hfill$\Box$

\subsection{Control of the remainder term.}\label{s7.3}

The control of the error terms in  Theorem \ref{t7.2}, that is to say,  points
  (\ref{7.4}), (\ref{7.6}) and (\ref{7.7}), are a consequence of the following  Theorem
together with the construction  in Proposition \ref{p7.3}.

 Set $L $ a selfadjoint  extension  in $H$ of the operator,
\be\label{7.3.1} L = {1\over i} {\partial \over \partial r } + {1\over i r},\ee
where $r = |k|$.

\begin{theo}\label{t7.7}  Let $U_h$ be an element of $D( H(h))$,
   satisfying,
\be\label{7.3.2} H(h) U_h = \lambda (h) U_h + R_h.\ee
Suppose that there are $C>0$ and $p\geq 1$ such that, for all $h$ in $(0, 1)$,
\be\label{7.3.3} \Vert U_h - ( \Psi_0 \otimes a_{\emptyset })  \Vert \leq  C h^{1/2}. \ee
Suppose also tat $U_h$ and $R_h$ are in the domain of ${\rm d}\Gamma (  L)  \otimes I$,
and that:
\be\label{7.3.4}  \Vert  U_h \Vert +  \Vert ( {\rm d}\Gamma (  L)  \otimes I )  U_h \Vert   \leq C,\qquad \Vert  R_h \Vert  +
\Vert ( {\rm d}\Gamma (  L)  \otimes I ) R_h \Vert   \leq C h^{p+2}.\ee
Then, for any sufficiently small $h$,
\be\label{7.3.5}|\lambda (h) - E_h | \leq 2 C h^{p+2}.\ee
  Set $\varphi_h$ a normalized eigenvector corresponding to the eigenvalue $E_h$,
  the infimum of the spectrum of  $H(h)$, satisfying  (\ref{7.2}).
  From (\ref{7.2}),  for $h$ small enough, one can choose $\rho_h >0$ and   $\theta _h $ such that the function $V_h$ defined by,
\be\label{7.3.6}V_h = U_h - \rho_h e^{i\theta _h } \varphi_h,\ee
satisfies,
\be\label{7.3.7} < V_h , \Psi_0 \otimes a_{\emptyset} > = 0.\ee
  The function $V_h$  then satisfies,
\be\label{7.3.8}\Vert  V_h \Vert \leq C h^{p+1}.\ee
We also have for  small enough $h$,
\be\label{7.3.9}  <  (N  \otimes I)V_h ,V_h  >\  \leq K h^{2p}.\ee
\end{theo}

{\it Proof of (\ref{7.3.5}).} In view of (\ref{7.2}) and (\ref{7.3.3}), one deduces,
$$ \Vert  \varphi_h -  U _h \Vert \
\leq C h^{1/2}.   $$
As a consequence, for small enough $h$,
$$ | < U_h , \varphi _h > | \geq 1/2.$$
By equaling the scalar products of the two hand sides of   (\ref{7.3.2}) with $\varphi_h$ which satisfies $H(h) \varphi _h = E_h \varphi _h $, one obtains,
$$ | E_h - \lambda (h) | \ |< U_h , \varphi_h > | \ \leq \Vert R_h \Vert
\leq C h^{p+2}.  $$
For $h$ small enough, inequality (\ref{7.3.5}) then follows.

\hfill$\Box$

Estimates  (\ref{7.3.8}) and  (\ref{7.3.9}) are a consequence of the two following Propositions. The first one is relying on
 a conjugated operator argument.

\begin{prop}\label{p7.8}
Let $V_h $ be an element of  $D( H(h))$
 and $f_h $ be  element of ${\cal H } _{ph} \otimes {\cal H } _{sp}$  satisfying,
\be\label{7.3.10}  ( H(h) - E_h) V_h = f_h,\ee
where  $E_h \leq - N |\beta |h$.
Suppose that $f_h$ belongs to the domain of ${\rm d}\Gamma (L)\otimes I $, where
$L $ a selfadjoint  extension in $H$ of the operator (\ref{7.3.1}). We suppose
also that (\ref{7.3.7}) is satisfied,  and that
\be\label{small-h} 16 h^{1/2}  \sum _{\lambda = 1}^N
\sum _{m=1}^3 | L A_{m} (x_{ \lambda}  ) | \leq 1 \hskip 2cm
4 \frac {h^{1/2}} {|\beta |} 2^{ 1 + (N/2)}\sum _{\lambda = 1}^N
\sum _{m=1}^3 | A_{m} (x_{ \lambda}  ) | \leq 1  .\ee
Then we have:
\be\label{ESTIM-1}  \Vert V_h \Vert \leq \frac {16} {h} \Vert ({\rm d}\Gamma (L)\otimes I ) f_h \Vert
\ + \ \frac {4} {|\beta| h} \Vert f_h \Vert   , \ee
\be\label{ESTIM-2}  < (N \otimes I) V_h , V_h > \leq \frac {300} {h^2}
\Vert ({\rm d}\Gamma (L)\otimes I ) f_h \Vert ^2 +  \frac {32} {h^2 |\beta |^2 }
\Vert f_h \Vert^2  .\ee
\end{prop}

{\it Proof. First step.}  When $M$ is the multiplication by  $|k|$, and and $L$ is defined in (\ref{7.3.1}), we have,
$$ [{\rm d}\Gamma (L) , {\rm d}\Gamma (M)] = {\rm d}\Gamma ([L, M]) = {1\over i} {\rm d}\Gamma (I)
= {1\over i} N.$$
If $V_h$ satisfies (\ref{7.3.10}), we have:
$$ {\rm Im} < ( H(h) - E_h) V_h , (d\Gamma (L) \otimes I ) V_h > =
      {\rm Im} <  f_h , (d\Gamma (L) \otimes I ) V_h >. $$
We use the notations (\ref{7.2.16}), (\ref{7.2.17}), (\ref{7.2.18}) for the
operator $H(h)$. Recalling that $H_{ph} = h d\Gamma (M)$, it follows from the
above commutator relation that:
$$ {\rm Im} < (H_{ph} \otimes I) V_h ,  (d\Gamma (L)  \otimes I) V_h >
= -\frac {h} {2} < (N \otimes I) V_h , V_h > $$
We have
$$ {\rm Im} <  ( ((I \otimes T_0 ) - E_h ) V_h  , (d\Gamma (L)  \otimes I) V_h >  = 0$$
Therefore:
$$ -\frac {h} {2} < (N \otimes I) V_h , V_h > + h^{3/2}  {\rm Im} <   K_{3/2} V_h , (d\Gamma (L)  \otimes I) V_h > = {\rm Im} < f_h , (d\Gamma (L)  \otimes I) V_h >$$
Setting $A_m ( x _{\lambda } ) = a_m ( x _{\lambda } ) + i b_m ( x _{\lambda } ) $,
and $L$ is defined in (\ref{7.3.1}), we have
classically:
$$ i [ {\rm d}\Gamma (L) , \Phi_S ( A_{m} (x_{ \lambda}  ) )  ] =
 \Phi_S ( i L  A_{m} (x_{ \lambda}  )  )
 $$
By (\ref{8bis}), we have:
$$ \Vert ( \Phi_S (i  L A_{m} (x_{ \lambda}  ) ) \otimes \sigma _m ^{[\lambda]} )   V_h \Vert^2 \leq
   |L A_{m} (x_{ \lambda}  ) |^2  \ \Big [
\Vert V_h \Vert ^2 + < (N \otimes I) V_h , V_h > \Big ] .$$
Therefore:
$$ \frac {h} {2} < (N \otimes I) V_h , V_h > \leq  \frac {h^{3/2} } {2} \Vert V_h \Vert
\ \Big [ \Vert V_h \Vert ^2 + < (N \otimes I) V_h , V_h > \Big ]^{1/2}
\sum _{\lambda = 1}^N \sum _{m=1}^3 |L A_{m} (x_{ \lambda}  ) | + ...$$
$$ ... + \Vert (d\Gamma (L)  \otimes I) f_h \Vert \ \Vert V_h \Vert  $$
If (\ref{small-h}) is satisfied, then we have:
\be\label{first-step}  < (N \otimes I) V_h , V_h > \ \ \leq \ \ \frac {4} {h}  \Vert ({\rm d}\Gamma (L)\otimes I)  f_h \Vert
 \ \Vert V_h \Vert
+ \frac {1} {16}   \Vert V_h \Vert ^2. \ee
{\it Second step.} Let us denote by $P_{\emptyset }$ the projection in
${\cal H}_{sp}$ on the vectorial line generated by  $a_{\emptyset}$ and by
$ P_{\emptyset }^{\perp}$ the projection on the orthogonal subspace. Also, $P_{\Omega }$ denotes the projection in ${\cal H}_{ph}$ on the vacuum $\Psi_0$ and $ P_{\Omega }^{\perp}$ stands for the projection on the orthogonal subspace.  By (\ref{7.3.10}), we have:
$$  < ( P_{\Omega } \otimes P_{\emptyset }^{\perp}  )(H(h) - E_h)  V_h , V_h > =
 < ( P_{\Omega } \otimes P_{\emptyset }^{\perp}  ) f_h , V_h > $$
 We have $ P_{\Omega }  H_{ph} = 0$. Note that $ P_{\emptyset }^{\perp} T_0  \geq |\beta | ( 1 - N) P_{\emptyset }^{\perp} $.
In particular, if  $E_h \leq - N |\beta |h$,
$$ < ( P_{\Omega } \otimes P_{\emptyset }^{\perp}  ) (I \otimes  hT_0 - E_h) )  \geq
h |\beta | < ( P_{\Omega } \otimes P_{\emptyset }^{\perp}  ) V_h , V_h > .  $$
Therefore:
$$ h |\beta | < ( P_{\Omega } \otimes P_{\emptyset }^{\perp}  ) V_h , V_h > \leq
h^{3/2} |  < ( P_{\Omega } \otimes P_{\emptyset }^{\perp}  ) K_{3/2} V_h , V_h > | +
| < ( P_{\Omega } \otimes P_{\emptyset }^{\perp}  ) f_h , V_h > | $$
We have:
$$  | < ( P_{\Omega } \otimes P_{\emptyset }^{\perp}  ) K_{3/2} V_h , V_h >|  \leq
\sum _{E \not = \emptyset }  | < V_h ,  K_{3/2}
( \Psi_0 \otimes a_E ) > | \  | < V_h , ( \Psi_0 \otimes a_E ) > | $$
$$ \leq M \Vert ( P_{\Omega } \otimes P_{\emptyset }^{\perp}  ) V_h\Vert
\ \Vert V_h \Vert $$
with:
$$ M =  \left [ \sum _{E \not = \emptyset } \Vert  K_{3/2} ( \Psi_0 \otimes a_E )
\Vert ^2 \right ]^{1/2} \leq 2^{N/2} \sum _{\lambda = 1}^N \sum _{m=1}^3
\Vert \Phi_S (A_m (x_{\lambda })) \Psi_0 \Vert $$
By (\ref{8bis})
$$ \Vert \Phi_S (A_m (x_{\lambda })) \Psi_0 \Vert  \leq 2 |A_m (x_{\lambda })|$$
Therefore, if (\ref{small-h}) is satisfied:
\be\label{sec-step}  \Vert ( P_{\Omega } \otimes P_{\emptyset }^{\perp}  ) V_h \Vert \leq
\frac {1} {16} \Vert V_h \Vert + \frac {1} {h |\beta |} \Vert f_h \Vert .  \ee
{\it Third step.} By the  condition (\ref{7.3.7}), we have:
$$ \Vert V_h \Vert \leq \Vert (P_{\Omega}^{\perp} \otimes I)  V_h \Vert +
   \Vert ( P_{\Omega } \otimes P_{\emptyset }^{\perp}  ) V_h \Vert $$ \hskip 4cm \ 
$$ \ \hskip 4cm  \leq \  < (N \otimes I) V_h , V_h >^{1/2} + \Vert ( P_{\Omega } \otimes P_{\emptyset }^{\perp}  ) V_h \Vert $$
and therefore (\ref{ESTIM-1}) follows from (\ref{first-step}) and (\ref{sec-step}), and (\ref{ESTIM-2}) follows from(\ref{ESTIM-1}) and (\ref{first-step}).

  {\it End of the proof of Theorem \ref{t7.7}.} Estimate (\ref{7.3.5}) is already proved. If $U_h$ satisfies (\ref{7.3.2}), then we have,
\be\label{f_h} H(h) U_h = E_h U_h + f_h,\qquad f_h = R_h + ( \lambda (h) - E_h) U_h.\ee
For any small enough $h$, we can choose  $\rho_h >0$ and   $\theta _h $ such that, the function $V_h$
defined by (\ref{7.3.6}) satisfies (\ref{7.3.7}) and also (\ref{7.3.10}). The functions $R_h$
and $R_h$, annd therefore $f_h$, are in the domain of ${\rm d} \Gamma (L) \otimes I$. Therefore, by
Proposition \ref{p7.8}, if the conditions (\ref{small-h}) are satisfied, then the estimates
(\ref{ESTIM-1}) and (\ref{ESTIM-2}) are satisfied. By (\ref{7.3.4}) and (\ref{7.3.5}), we have:
 $$\Vert  f_h \Vert  +
\Vert ( {\rm d}\Gamma (  L)  \otimes I ) f_h \Vert   \leq C h^{p+2}.$$
The estimates (\ref{7.3.8}) and (\ref{7.3.9}) follow from (\ref{ESTIM-1}), (\ref{ESTIM-2})
and the above inequality.

\hfill $\Box$

{\it End of the proof of Theorem \ref{t7.2}.} We apply Theorem \ref{t7.7} with the
elements $U(h) = U^{(2p+1)} (h)$ and $R(h) = R ^{(2p+1)} (h)$ and with the real number $\lambda (h) = \lambda ^{(p+1)} (h)$ of Proposition \ref{p7.3}. These elements satisfy (\ref{7.3.2}) from (\ref{7.2}).
The condition  (\ref{7.3.3}) comes from the fact that $u_0$ is defined in (\ref{7.3}) and that the other
$u_j$ are independent on $h$. The assumption (\ref{7.3.4}) comes from the fact that the $u_j$ and the
$ f_{2p +1 }^{ (p + 2 + (k/2)) } $  of Proposition \ref{p7.3} are finite sums of terms all belonging to the spaces  ${\cal F } _m (\rho)$. Note that ${\cal F } _m (\rho)$ is invariant by the operator ${\rm d}\Gamma (L)$ where $L$
is defined in (\ref{7.3.1}).  The hypotheses of Theorem
\ref{t7.7} are satisfied. Inequality (\ref{7.4}) follows from (\ref{7.3.5}). Inequality (\ref{7.6})
is a consequence of (\ref{7.3.8}) and inequality (\ref{7.7}) comes from (\ref{7.3.9}).

{\it Proof of  Theorem \ref{t1.2}.} The point i) follows from inequality (\ref{7.4}) of theorem 
 \ref{t7.2}. For the point ii), without any hypothesis on $\chi (0)$,
the elements $u_0$ and $\lambda _1$ are defined by (\ref{7.3}), and $u_1$ and $\lambda _2$
are constructed in  Proposition  \ref{p7.5}.  Let us prove that the construction of $u_2$
 is also possible without any hypothesis on $\chi (0)$. Let us now define $u_2$ which has to satisfy
 (\ref{7.2.21}), that is to say, taking into account the choice of  $\lambda _2$ in (\ref{7.2.33}),
 $$ ( K_1 - \lambda _1) u_2 = - ( I- \Pi) K_{3/2} u_1,$$
 where $\Pi$ is the orthogonal  projection on the vectorial line generated by
  $u_0 = \Psi_0 \otimes a_{\emptyset}$. Since $u_1$ is defined in (\ref{7.2.31}), we can write
$$  K_{3/2} u_1 = \sum _{ E \subset \{ 1 ,\dots, N \} }
(f_E + g_E) \otimes a_E,$$
where the $f_E$ belong to ${\cal F}_0 $ and the $g_E$ lie in ${\cal F}_2 $.
We have $\Pi  K_{3/2} u_1 = f _{\emptyset } \otimes a _{\emptyset } $. Consequently,
\be\label{u_2}  u_2 = \sum _{ E \subset \{ 1 , \dots, N \} }
(u_E + v_E) \otimes a_E,\ee
where the $u_E$ are in ${\cal F}_0 $ and the $v_E$ in ${\cal F}_2 $.
We shall have $u _{\emptyset } = 0$. If $E$ is non empty then we shall obtain, according to
Lemma 7.4, $u_E = f_E / (2 |\beta| |E| )$. The elements $g_E$ and $v_E$ being identified with symmetric functions on $\R^3 \times
\R^3$ and taking vector values, we have,
\be\label{v_E} v_E ( k_1 , k_2) = { g_E ( k_1 , k_2) \over |k_1| + |k_2| + 2 |\beta | |E|}.  \ee
 According to (\ref{7.2.29})-(\ref{7.2.32}), the functions $g_{E} $ are linear combinations of products of the form
 $ (a_m (x_{\lambda }) + i b_m (x_{\lambda }) ) ( k_1)
u_{\mu } (k_2)$, of  products where the second factor is $u_{\emptyset } (k_2)$, and of products where the factors are exchanged. The $u_{\mu }$ and $u_{\emptyset }$
are rapidly decreasing at infinity and, when $k$ tends to $0$,
$u_{\mu } (k) = {\cal O } (|k|^{1/2} )$ and $u_{\emptyset } (k) = {\cal O } (|k|^{-1/2} )$.
Concerning elements $ (a_m (x_{\lambda }) + i b_m (x_{\lambda }) ) ( k) $, they
are an ${\cal O } (|k|^{1/2} )$ when $k$ tends to $0$. Consequently,
equalities (\ref{v_E}) therefore define elements $v_E$ of  ${\cal F}_2 $,
 and (\ref{u_2}) indeed defines an element $u_2$ of $( {\cal F}_0 \oplus {\cal F}_2) \otimes
{\cal H} _{sp}$  (Without vanishing assumptions on $\chi$ at the origin, it does not seem possible to further follow  the expansion).  The element $K_{3/2} u_2 $ is well defined since $K_{3/2}$ is
continuous from  $( {\cal F}_0 \oplus {\cal F}_2) \otimes
{\cal H} _{sp}$ into $( {\cal F}_1 \oplus {\cal F}_3) \otimes
{\cal H} _{sp}$.  We then have,
 $$ ( H(h) - \lambda _1 h -  \lambda _2 h^2) \ ( u_0 + h^{1/2} u_1
 + h u_2) = R^{(2)} (h) = h^{5/2} ( K_{3/2} u_2 - \lambda _2 u_1 ) - h^3 \lambda _2 u_0.$$
Thus, we have $\Vert  R^{(2)} (h) \Vert \leq C h^{5/2}$. Taking the scalar products of both sides with $\varphi_h$ satisfying (\ref{7.2}) and $( H(h) - E_h ) \varphi_h = 0$, one therefore obtains estimate (\ref{1.B}).

\hfill$\Box$

{\it Proof of Theorem \ref{t1.3}.} Without any hypothesis on $\chi (0)$, we
determined $u_0$ and $\lambda _1$ from (\ref{7.3}), and $u_1$ according to Proposition  \ref{p7.5}.
One can choose $\rho_h$ and $\theta _h$ such that the following function,
$$ V_h = u_0 + h^{1/2} u_1 - \rho_h e^{i \theta _h } \varphi_h $$
satisfies (\ref{7.3.7}).  According to Proposition  \ref{p7.6}, we have,
$$ {\bf B}^{class}   (x) = < ({\bf B}  (x)\otimes I)  ( u_0 + h^{1/2} u_{1} ), ( u_0 + h^{1/2} u_{1} ) >. $$
Thus,
\be\label{etoile} \left |  B_m^{class}   (x) - \rho_h ^2 < B_m(x) \varphi_h , \varphi_h > \right |
\leq \Vert B_m(x) V_h  \Vert \ \Big ( 2 \rho_h + \Vert V_h \Vert ^2 \Big ). \ee
By the construction in Proposition  \ref{p7.5}, especially (\ref{7.2.19}) and (\ref{7.2.20}),
 the function $V_h$ defined above satisfies
(\ref{7.3.10}) with,
$$ f_h = h^2 K_{3/2} u_1 + ( \lambda _1 h - E_h) ( u_0 + h^{1/2} u_1). $$
According to point $(ii)$ of Theorem \ref{t1.2}, we have $| \lambda _1 h - E_h | \leq C h^2$.
Now, if $\chi(0) = 0$, let us prove that  $u_1$ and   $ K_{3/2} u_1 $ lie in the domain
of the operator ${\rm d}\Gamma (L) \otimes I$.
One follows  the construction of $u_1$ given in (\ref{7.2.31}). The elements
$u_{\emptyset }$ and $u_{\lambda }$ of ${\cal F}_1$ defined in  (\ref{7.2.32}) can be identified to elements of $H_{\C}$ and then to functions on $\R^3$ and taking values  in $\C^3$.
In general, $u_{\emptyset } (k) = {\cal O} ( |k|^{-1/2})$ when $k$ tends to $0$ and this
function is not belonging to the domain of the operator $L$ defined in (\ref{7.3.1}). However, if $\chi (0) = 0$, we have $u_{\emptyset } (k) = {\cal O} ( |k|^{1/2})$ when $k$ tends to $0$
 and $u_{\emptyset } $ lies in $D(L)$. This also holds true for $u_{\lambda}$. Consequently,  $u_1$ lies in the domain of the operator ${\rm d}\Gamma (L) \otimes I$. This is also valid for $ K_{3/2} u_1 $. Therefore, $f_h$ belongs to the domain of
${\rm d}\Gamma (L) \otimes I$ and
$$\Vert  f_h \Vert  +  \Vert ({\rm d}\Gamma (L) \otimes I) f_h \Vert \leq C h^2. $$
Then, we can apply  Proposition  \ref{p7.8}. This enables to write, for $h$ small enough,
$$ \Vert V_h \Vert \leq  \leq C h  ,\qquad < (N \otimes I) V_h , V_h > \ \ \leq  C h^{2} .$$
According to (\ref{8bis}),
$$ \Vert B_m(x) V_h  \Vert  \leq  C h^{1/2}  \Vert  V_h  \Vert + Ch^{1/2}  < (N\otimes I)  V_h , V_h  > ^{1/2}.$$
Thus, $\Vert  B_m(x) V_h  \Vert  \leq  C h^{3/2}$. Condition (\ref{7.3.7}),
Definition (\ref{7.3}) of $u_0$ and the property (\ref{7.2}) for $\varphi_h$
imply  that $ | 1 - \rho_h e^{ i \theta _h} | \leq C h^{1/2}$.
The right hand side of (\ref{etoile}) is then  ${\cal O } (h^{3/2}) $.
Since ${\bf B_m}^{class}   (x) $ is an ${\cal O } (h)$, the above equality (\ref{etoile})
shows that $< B_m(x) \varphi_h , \varphi_h > $ is  also an ${\cal O } (h)$.
Consequently,
$$ ( \rho_h ^2 - 1 ) \ < B_m(x) \varphi_h , \varphi_h > =  {\cal O } (h^{3/2} ). $$
Theorem 1.3 then follows.

\hfill$\Box$

\medskip

laurent.amour@univ-reims.fr\newline
LMR EA 4535 and FR CNRS 3399, Universit\'e de Reims Champagne-Ardenne,
 Moulin de la Housse, BP 1039,
 51687 REIMS Cedex 2, France.

jean.nourrigat@univ-reims.fr\newline
LMR EA 4535 and FR CNRS 3399, Universit\'e de Reims Champagne-Ardenne,
 Moulin de la Housse, BP 1039,
 51687 REIMS Cedex 2, France.


\begin{thebibliography}{99}





\bibitem{A-J-N} L. Amour, L. Jager, J. Nourrigat, {\it On bounded Weyl
pseudodifferential operators in Wiener spaces,}  Journal of Functional Analysis 269 (2015),
 pp. 2747-2812.




\bibitem{A-L-N-1} L. Amour, R. Lascar, J. Nourrigat, {\it Beals characterization of pseudodifferential
 operators in Wiener spaces}, Appl. Math. Res. Express (2016).




\bibitem{A-L-N-2} L. Amour, R. Lascar, J. Nourrigat, {\it Weyl calculus in QED I. The unitary group,}
 preprint, arXiv:1510.05293, october 2015.


\bibitem{A-H}
A. Arai, M. Hirokawa, {\it On the existence and uniqueness of ground states of a generalized spin-boson model}, J. Funct. Anal. 151 (1997), no. 2, 455--503.

\bibitem{B-F-S} V. Bach, J. Fr\"ohlich, I. M. Sigal, {\it Quantum electrodynamics of confined nonrelativistic particles,} Adv. Math. 137 (1998), no. 2, 299--395.


\bibitem{BL} F. Bloch, {\it Nuclear Induction}, Physical Review {\bf 70} 460-473, (1946).



\bibitem{D-G} J. Derezi\'nski, C.  G\'erard,  {\it Asymptotic completeness in quantum field theory.
 Massive Pauli-Fierz Hamiltonians. }
Rev. Math. Phys. 11 (1999), no. 4, 383-450.

\bibitem{G} C. G{\'e}rard,
{\it On the existence of ground states for massless Pauli-Fierz Hamiltonians.},
Ann. Henri Poincar\'e
{\bf 1}
(2000), 443--459.




\bibitem{H-H-L}  M. Hirokawa, F. Hiroshima, J. L{\H{o}}rinczi, {\it  Spin-boson model through a Poisson-driven stochastic process},
Math. Z. {\bf 277}
(2014),
1165--1198.

\bibitem{H-S} M. H\"ubner, H. Spohn, {\it Spectral properties of the
spin-boson Hamiltonian.}, Annales de l'I. H. P., Section A ,
tome 62, {\bf 3} (1995), 289-323.


\bibitem{L-L} E. Lieb, M. Loss, {\it  A note on polarization vectors in quantum electrodynamics.}
  Comm. Math. Phys. {\bf 252} (2004), no. 1-3, 477-483.




\bibitem{RE-SI} M. Reed, B. Simon, {\it Methods of modern
mathematical physics,}  Vol II, Fourier Analysis, selfadjointness,
Academic Press, 1975.



\bibitem{REU} F. A. Reuse, {\it Electrodynamique et Optique
Quantiques,}  Presses Polytechniques et Universitaires Romandes,
Lausanne, 2007.


\end{thebibliography}
\end{document}